\documentclass[letterpaper,twocolumn,showpacs,prl,superscriptaddress,floatfix]{revtex4}
\usepackage[german,swedish,english]{babel}
\usepackage[dvips]{graphicx}
\usepackage{amssymb}
\usepackage{amsmath}
\usepackage{color}

\begin{document}

\date{\today}

\title{Fermi-surface induced modulation in an optimally doped YBCO superconductor}
\author{Xuerong Liu}
\affiliation{Physics Department, University of California, San Diego, 9500 Gilman Drive, La Jolla, California 92093\\}
\author{Zahirul Islam}
\affiliation{Advanced Photon Source,Argonne National Laboratory, 9700 S. Cass Ave., Argonne IL 60439\\}
\author{Sunil K. Sinha}
\affiliation{Physics Department, University of California, San Diego, 9500 Gilman Drive, La Jolla, California 92093\\}
\author{Simon C. Moss}
\affiliation{Department of Physics and Texas Center for Superconductivity, University of Houston, Houston, Texas 77204\\}
\author{Robert J. McQueeney}
\affiliation{Department of Physics and Astronomy, Iowa State University, Ames, IA 50011\\}
\author{Jonathan C. Lang}
\affiliation{Advanced Photon Source,Argonne National Laboratory, 9700 S. Cass Ave., Argonne IL 60439\\}
\author{Ulrich Welp}
\affiliation{Materials Science Division, Argonne National Laboratory, Argonne, Illinois 60439\\}

\begin{abstract}
We have observed a Fermi-surface (FS) induced lattice modulation in a YBCO superconductor with a wavevector along CuO chains, {\it i.e.} ${\bf q}_1$=(0,$\delta$,0). The value of $\delta\sim0.21$ is twice the Fermi wavevector ($2{\bf k}_F$) along {\bf b*} connecting nearly nested FS `ridges'. The ${\bf q}_1$ modulation exists only within O-vacancy-ordered islands (characterized by ${\bf q}_0$=$\left(\frac14,0,0\right))$ and persists well above and below $T_c$. Our results are consistent with the presence of a FS-induced charge-density wave.
\end{abstract}


\maketitle

\begin{figure}[h]
\includegraphics[width=0.29\textwidth]{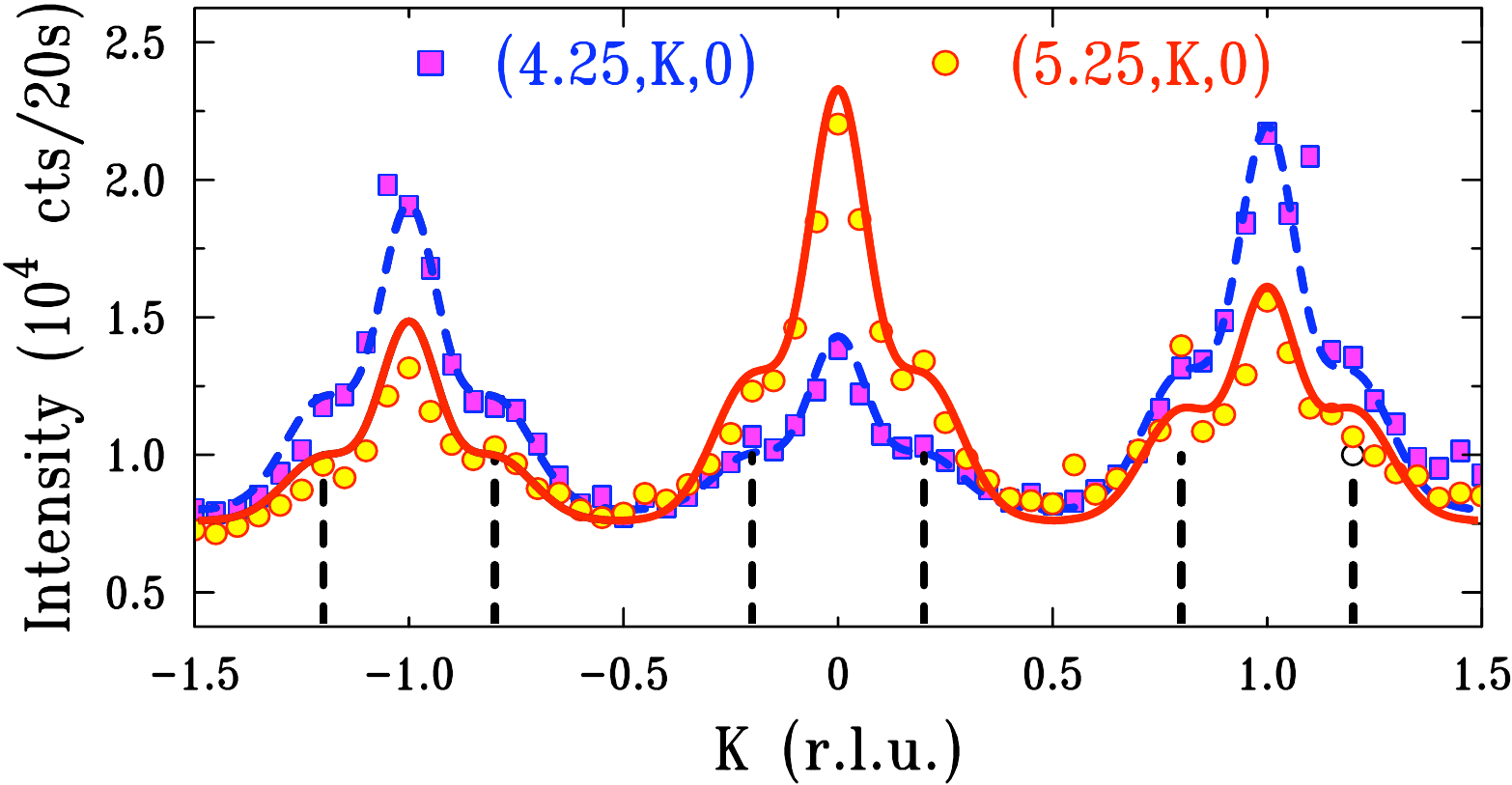}\hspace{0.02in}\includegraphics[width=0.18\textwidth]{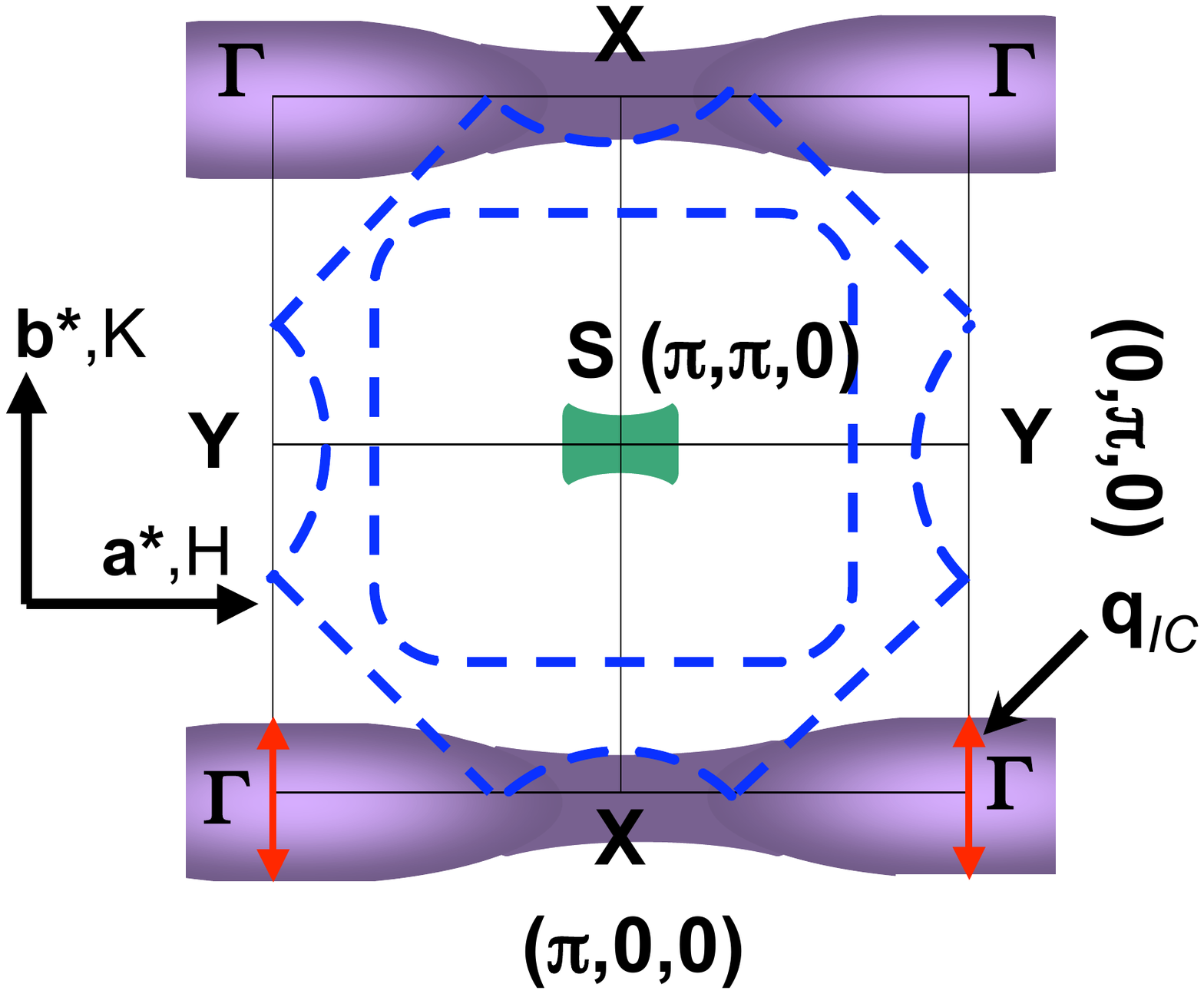}
\caption{Left: {\bf K} scans (T$\sim$7K) showing shoulders on both sides of the central ${\bf q}_0$ satellite peaks (lines are from Gaussian profile as a guide to the eye). Vertical dashes indicate the shoulders displaced along K by $\pm 0.21$ from their respective ${\bf q}_0$ satellite. Right: A schematic showing near nesting of the `ridges' (purple) with nesting vector along {\bf b*} (red)\cite{ACAR,PhoEmiss,ACAR-2}.}
\label{fig-lines}
\end{figure}

YBa$_2$Cu$_3$O$_{7-x}$ (YBCO) compounds possess a unique structure where one-dimensional CuO chains are interleaved with two-dimensional CuO$_2$ bi-layers, from which superconductivity appears to emerge.  These chains not only act as the charge reservoirs for the CuO$_2$ planes, but also provide the opportunity for the interplay of electronic properties of systems with different dimensionality in close proximity to each other. Various studies\cite{resistivity,infrared,thermal} have revealed that the chains exert profound influence on the transport properties in both normal and superconducting states. Local density approximation(LDA) calculations\cite{bandcalcu} have indicated sections of Fermi surface (FS) that are open electron ``ridges''  ($\parallel \Gamma X$) orthogonal to {\bf b*} ({\it i.e.} normal to the CuO chain direction, see Fig.\ \ref{fig-lines}).  These ridges are due to CuO-chain derived electronic states, and have been confirmed by positron annihilation\cite{ACAR} and photoemission\cite{PhoEmiss} studies. Intriguingly, near the zone center these ridges remain as parallel sheets\cite{ACAR-2} along {\bf c*} forming a near nesting condition with an incommensurate wavevector ${\bf q}_\text{IC}=(0, 2k_F\approx 0.22, 0)$ (in reduced lattice units; ${\bf k}_F$ is the Fermi wavevector), suggesting that the CuO chains may be on the verge of an electronic instability ({\it e.g.} a charge-density wave, CDW). 

Although local probes such as NMR/NQR measurements\cite{NMRNQR} have reported on the formation of a CDW on the CuO chains, the results of different groups have been at odds with each other. Observations of electronic modulations of the chains using STM\cite{STM-Edwards,STM-Maki} have seemed promising because the observed periodicity was consistent with the expected ${\bf q}_\text{IC}$ for an optimally doped YBCO, but doubts have been raised\cite{STM-Derro} on the CDW interpretation due to the dispersive behavior of the modulation wavevector on energy. Furthermore, it remains unclear whether a {\it static} CDW on the 1D CuO chains as was suggested by STM work\cite{STM-Edwards,STM-Maki} forms in the bulk or not. Inelastic neutron scattering (INS) measurements\cite{INS} have also been used to infer the presence of charge fluctuations with wavevector of $(0,\sim0.23,0)$. These measurements, however, could not distinguish between {\bf a} and {\bf b} directions due to the twinned samples used in the experiment, adding an additional directional ambiguity.

In this Letter, we present x-ray scattering observations of lattice modulations with a wavevector ${\bf q}_1=\left(0, \sim0.21, 0\right)$ along the CuO chain direction. Atomic displacements associated with them are parallel to the orthorhombic {\bf a} axis. They co-exist with the O-ordered ORTHO-IV phase\cite{DdF02,A-modulation} islands characterized by  ${\bf q}_0=\left(\pm\frac14, 0, 0\right)$, rather than forming homogeneously throughout the crystal. In reciprocal-space scans along {\bf b*} ({\bf K} scans, Fig.\ \ref{fig-lines}), the ${\bf q}_1$ super-lattice peaks manifest themselves as shoulders on either side of the central ${\bf q}_0$ satellites.  Below we summarize our previous finding, then show the coupled nature of these two modulations. A quantitative modeling of the x-ray diffuse scattering pattern in the H-K plane is used to establish the source of the shoulders.
 
\begin{figure}[h]
\includegraphics[width=0.35\textwidth]{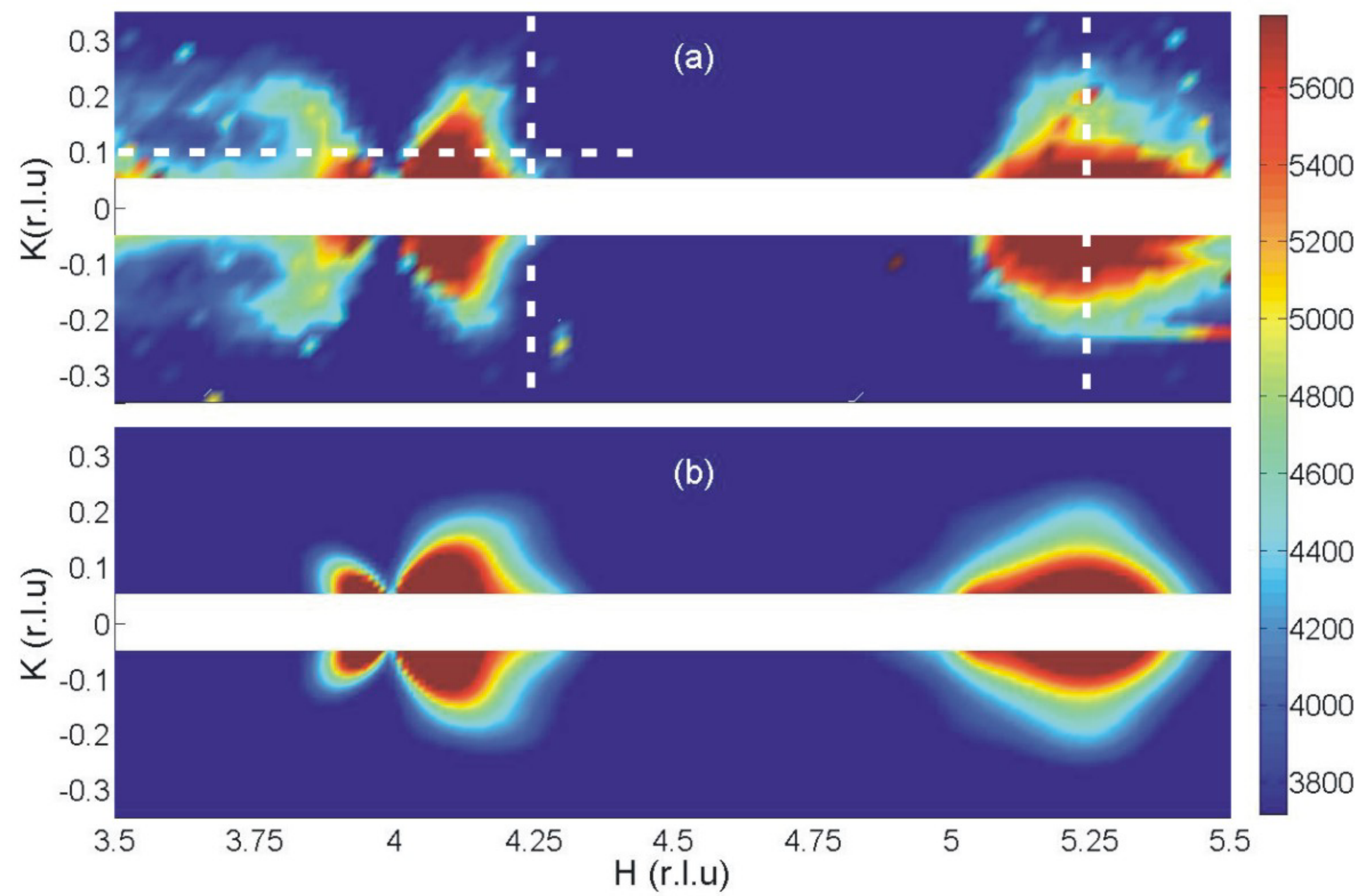}
\caption{(a): Colormap of the measured diffuse scattering around (4, 0, 0) and (5, 0, 0) in H-K plane. (b) Model calculations of the 2D scattering pattern. Asymmetric HDS around (4, 0, 0),  ${\bf q}_0$ satellite, and ${\bf q}_1$ shoulders are all reproduced. The dashed lines indicate the line cuts presented in Fig.\ \ref{fig:fit}.} 
\label{fig:data}
\end{figure}

\begin{figure}
\includegraphics[width=0.21\textwidth]{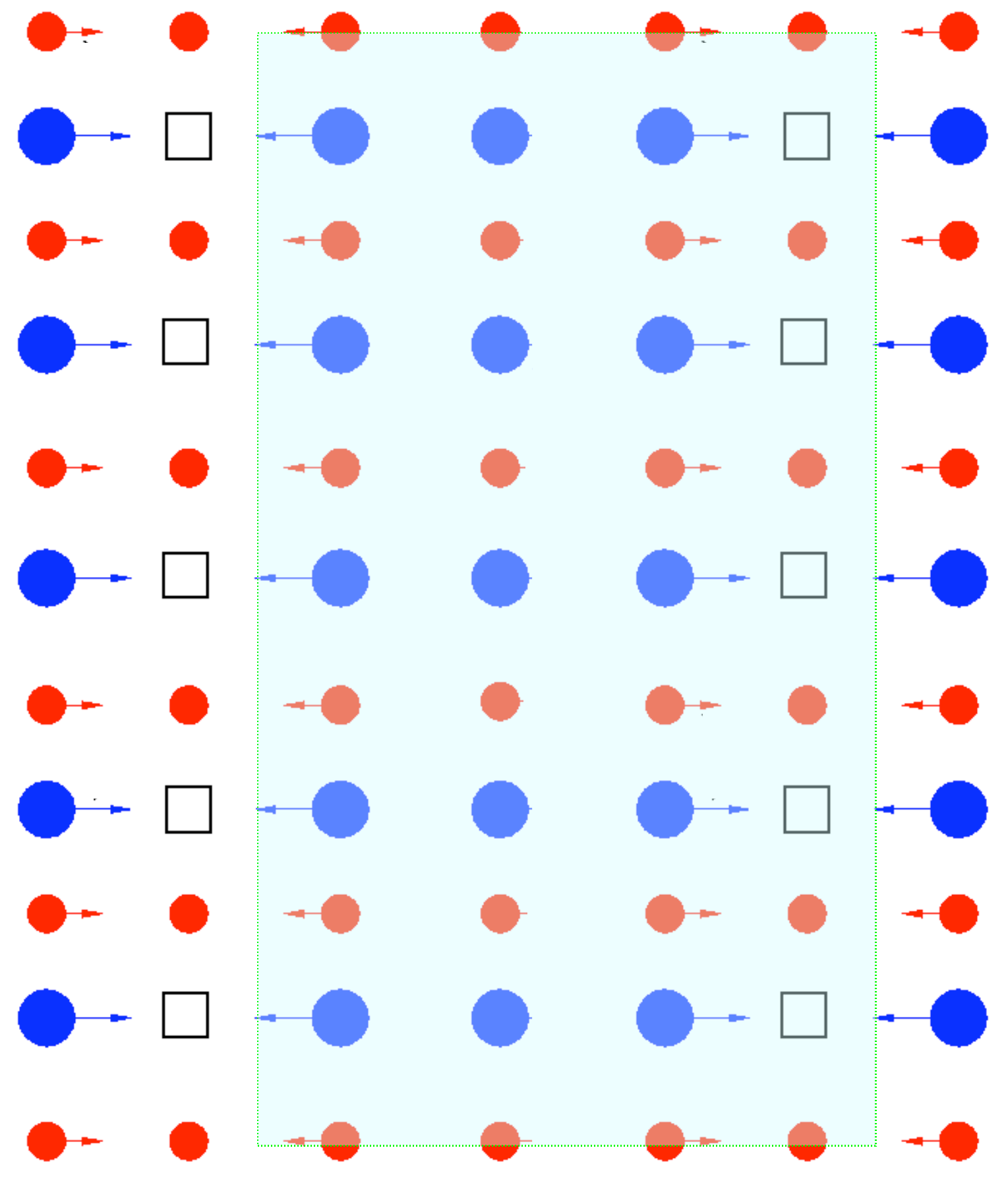}\hspace{0.15in}\includegraphics[width=0.25\textwidth]{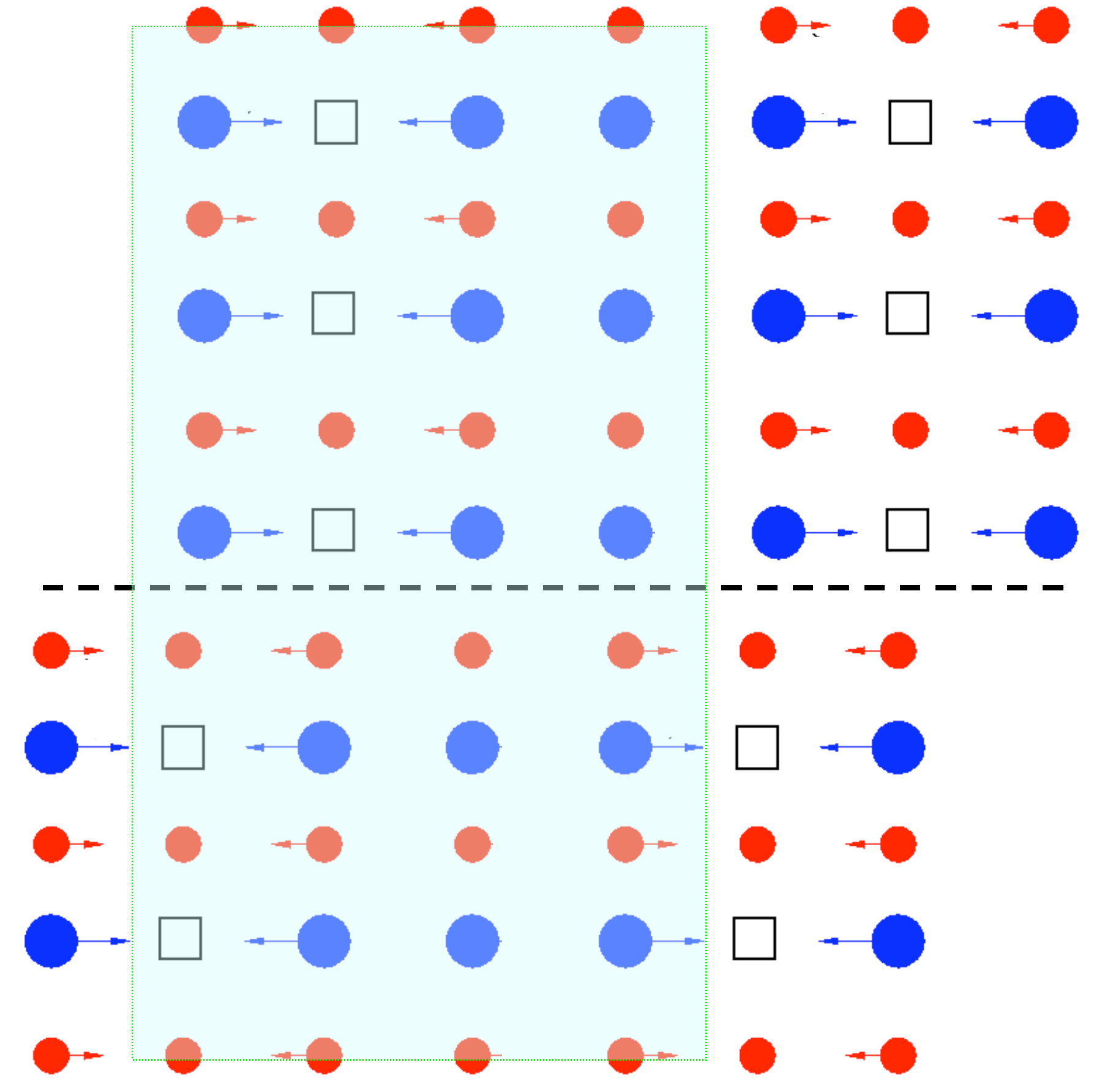}
\caption{Left: ORTHO-IV displacement pattern ($ab$-plane) for the chains\cite{A-modulation}. ${\bf q}_1$ further modulates amplitude of these displacements sinusoidally. Right: One possible AP domain with no modulations of displacement amplitudes. Small circles (red): Cu; Big circles (blue): Oxygen; Squares: Vacancy; Dashed line: Domain boundary; Shaded area: 4X5 super-cell.}
\label{APD}
\end{figure}

A high-quality {\it detwinned} crystal of optimally doped YBCO$_{7-x}$  ($x=0.08$ and $T_c$ = 91.5K) was used in our study, which eliminates any directional ambiguity. We note that the presence of ${\bf q_0}$ and ${\bf q}_1$ modulations was verified in a {\it twinned} crystal as well which will be published elsewhere.  High energy (E = 36keV) x-ray diffuse scattering experiments were performed on the X-Ray Operations and Research 4-ID-D beam line at the Advanced Photon Source (APS). The choice of high x-ray energy makes this a truly {\it bulk}-sensitive study\cite{A-modulation}. At this composition, first principles band calculations\cite{DdF02} have shown that the oxygen vacancies, located in the chains, tend to cluster together to form a minority phase with a 4-unit-cell superstructure, corresponding to a sequence of three O-full and one O-vacant chains (ORTHO-IV phase denoted by $<1110>$, see Fig.\ \ref{APD}), respectively, the formation of which was confirmed in a previous X-ray scattering study\cite{A-modulation}. That study revealed that, in the ORTHO-IV phase islands, coupled displacements of Cu, Ba, and O are induced, respectively, in the {\bf a}-{\bf c} plane with anisotropic correlation lengths along each of the three crystallographic axes, producing broad satellite peaks centered at ${\bf q}_0=\left(\pm\frac14, 0, 0\right)$ on both sides of Bragg points. Embedded in the average orthorhombic crystal, the ORTHO-IV phase islands induce long-range strain in the lattice which gives rise to the Huang diffuse scattering (HDS)\cite{ybco-hds} around Bragg peaks (Fig.\ \ref{fig:data}(a)). 

If lattice modulations associated with CDWs with propagating vector ${\bf q}_\text{IC}$ are present throughout the whole crystal, they should be observed by X-ray scattering as peaks (or rods) centered at $(m, n, L)\pm {\bf q}_{\text{IC}}$ ($m$ and $n$ are integers) in the H-K plane of the reciprocal space, depending on their inter-chain correlations. A thorough investigation in the vicinity of $(m, n, L)\pm {\bf q}_{\text{IC}}$ for different integer $m$ and $n$ values showed no indication of such modulations along {\bf b*} direction as claimed by the INS measurements\cite{INS}. Instead, clear shoulders were observed at ${\bf q}_0\pm{\bf q}_1$, ${\bf q}_1=(0, \delta, 0)$ with $\delta\sim{0.2}$, as shown in Fig.\ \ref{fig-lines}.  It is interesting to notice that {\bf K}-scans through different ${\bf q}_0$ peaks show a definite correlation between the intensities of the shoulder peaks with the center ORTHO-IV peaks, both having the same correlation lengths and obeying the same parity relations\cite{A-modulation}. The parity relations impose stringent phase correlations between atomic displacements such as those in the ORTHO-IV model\cite{A-modulation}.

Furthermore, careful measurements of the ratio of the satellite and the shoulder intensities as a function of increasing temperature found it to be constant in the range of 7-300K.  As pointed out in Ref.\ \cite{A-modulation}, the temperature dependence of the ${\bf q}_0$ peaks distinguishes the ORTHO-IV phase islands from the average lattice which has less rapidly varying Debye-Waller factors. The fact that the intensities of the shoulders obey the same temperature dependence as the center ${\bf q}_0$ peak strongly indicates that they are intimately related. Therefore, the ${\bf q}_1$ modulation must coexist with ORTHO-IV islands rather than forming homogeneously throughout the whole crystal.

Although a commensurate $\left(\pm\frac14,\pm\frac14,0\right)$-type superstructure in YBCO had been reported in the early literature\cite{quarter}, it was ruled out to be due to O-vacancy ordering in later studies\cite{Krekels}. Such superstuctures should give rise to a quartet of $\left(\pm\frac14,\pm\frac14,0\right)$ satellites of Bragg peaks, which is not observed here, suggesting that they were peculiar to the samples used in the electron diffraction work, and are not a general property of the bulk.

In order to identify the origin of the shoulders as shown in Fig.\ \ref{fig-lines}, two-dimensional scans around several Bragg peaks were performed and the diffuse scattering intensity pattern around (4, 0, 0) and (5, 0, 0) is shown in Fig.\ \ref{fig:data}(a). The salient feature of this pattern is the profound difference in the symmetry of scattering around the two Bragg peaks. While the dominating feature around the (4, 0, 0) Bragg peak is a strong nearly fourfold `bow-tie'-shape HDS pattern albeit asymmetric, it is entirely missing around (5, 0, 0). Rather, there is a strong `elliptical' pattern centered at the (5.25, 0, 0) ORTHO-IV satellite with additional scattering bulging out  along the K axis, which appears as shoulders in {\bf K} scan through (5.25, 0, 0). The intensity near (3.50, 0, 0) is from the tail of the strong ${\bf q}_0$-satellite at (3.25, 0, 0). In particular, the intensity asymmetry about the K axis through (4, 0, 0) is due to the interference scattering (IFS) from the ORTHO-IV phase islands and the lattice-strain fields, respectively. Such interference effects have been known in the study of disorder and extensively discussed in Refs.\cite{HDS}.

\begin{figure}
\includegraphics[width=0.40\textwidth]{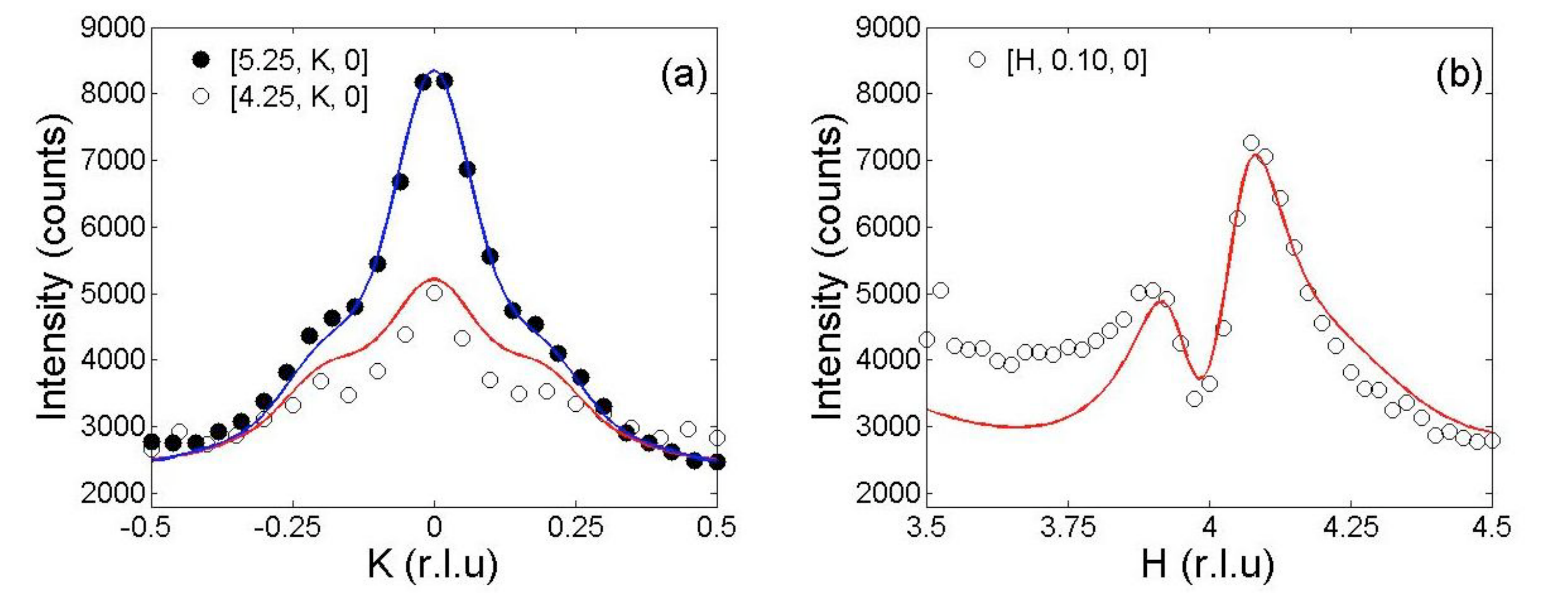}
\caption{Solid lines are the fitting results. $\delta$ and $d$ are determined to be 0.21 and 0.9, respectively. The ${\bf K}$ scan in (a) across (4.25, 0, 0) is over estimated by $\sim 13\%$ mainly due to the IFS contribution and the lower side of the ${\bf H}$ scan in (b) is pushed up by the tail of strong (3.25, 0, 0) satellite peak.}
\label{fig:fit}
\end{figure}

The presence of both HDS and IFS has to be taken into account in discussing the origin of the shoulders. The diffuse scattering pattern around (4, 0, 0) shown in Fig.\ \ref{fig:data}(a) suggests that the shoulders in the {\bf K}-scan through (4.25, 0, 0) in Fig.\ \ref{fig:fit}(a) might be due to the lobes of HDS, extending roughly along [1 1 0] directions in H-K plane. But this scenario is unlikely to account for the strong shoulders of the (5.25, 0, 0) satellite given the absence of similar HDS around (5, 0, 0) Bragg peak. 

The absence of HDS around (5, 0, 0) can be understood if one examines the general nature of HDS\cite{HDS}. As a long-range elastic distortion of the lattice, the strain fields give rise to HDS in the vicinity of a given Bragg point in proportion to the square of the unit cell structure factor modulus, 
$\left|F({\bf G})\right|^2=|\sum_k f_ke^{-W_k(Q)}e^{i{\bf G}\cdot{\bf R_k}}|^2$,
where ${\bf G}$ is the Bragg point\cite{HDS}. For YBCO$_{6.92}$, ${|\frac{F([500])}{F([400])}|}^2\thickapprox 0.02$. With this ratio, we estimate the HDS at (5.25, 0.2, 0) to be about $\sim20$ counts above background for the {\bf K}-scan in Fig.\ \ref{fig:fit}, assuming the shoulder at (4.25, 0.2, 0) is from HDS. The tail of the ${\bf q}_0$ peak is estimated to give $\sim20$ counts at (5.25, 0.2, 0) from Gaussian fitting. As a result, the interference term, IFS, can be at most $\sim40$ counts at (5.25, 0.2, 0). Both the HDS and IFS are negligible compared with the observed strong (5.25, $\delta$, 0) shoulders which are about $\sim2000$ counts. Thus, neither HDS nor IFS can account for the shoulders. They are the finger print of a lattice modulation along the {\bf b} axis.

We first discuss the X-ray diffuse scattering without considering the {\bf b}-direction modulation. With the presence of the ORTHO-IV phase islands and their surrounding strained lattice, the total diffuse scattering intensity, $I_{\mathrm{diffuse}}({\bf Q})$, can be written as in Eqn. (1). A linear approximation is made assuming that the lattice distortions due to the strain are small.
\begin{eqnarray}
I_{\mathrm{diffuse}}({\bf Q})\propto\!\!&<&\!\!\!|\sum_{jl^{'}k}f_k(Q)e^{-W_k(Q)}e^{-i{\bf Q}\cdot {\bf R}_{jl^{'}k}}(e^{-i{\bf Q}\cdot {\bf u}_{jl^{'}k}}\!-\!1)\nonumber \\
[0.5pt]&+&\!\!\!\!\sum_{lk}i{\bf Q}\cdot {\bf v}_{lk}f_k(Q)e^{-W_k(Q)}e^{-i{\bf Q}\cdot {\bf R}_{lk}}|^2\!\!>
\end{eqnarray}
$j$ is the index for the 4a x 1b x 1c supercell\cite{A-modulation} of an ORTHO-IV minority phase island. ${\bf u}_{jl^{'}k}$ is the lattice distortion vector of the $k$-th atom of the $l^{'}$-th unit cell inside the $j$-th supercell. Outside of the island, ${\bf u}_{jl^{'}k}$ is zero and the first term vanishes. ${\bf v}_{lk}$ is the lattice distortion of the $k$-th atom of the $l$-th unit cell due to the strain and the summation of the second term is over the whole crystal. $f_k(Q)$ and $e^{-W_k(Q)}$ are the atomic scattering factors and Debye-Waller factors (both dynamic and static), respectively. The modulus squared of the first term gives rise to the ORTHO-IV phase satellites, the modulus squared of the second term describes the HDS from lattice strain and the interference of these two terms gives rise to IFS. The additional thermal diffuse scattering was calculated using the shell model by fitting to the measured phonon dispersion curves\cite{phonon-disp} and subtracted from data except from those in Fig.\ \ref{fig-lines}.

The scattering from different contributions dominate different regions in reciprocal space. For example, the HDS has a minimum at all ${\bf q}_0$ positions and gives negligible contributions around Bragg peaks with odd parity. This allows us to decouple most of the parameters and fit them separately. In particular, in Ref.\cite{A-modulation}, ${\bf u}_{jl^{'}k}$ has been fitted to the integrated intensities of a set of ${\bf G \pm q_0}$ peaks alone (${\bf G}$ is a reciprocal lattice vector) which were extracted by a careful line fitting procedure, and an ORTHO-IV displacement model was constructed.

Now, we need to incorporate the {\bf b}-direction modulation into Eqn.(1).  As discussed above, the new modulation does not form homogeneously throughout the whole crystal. Rather, a ${\bf q}_1=(0,\delta,0)$-type wavevector acts as a {\bf b}-direction periodic modulation of the displacements of all the atoms involved in the ORTHO-IV phase\cite{A-modulation}. We incorporate this new modulation with a simple extension of the ORTHO-IV displacement pattern by replacing ${\bf u}'s$ in Eqn. (1) with ${\bf t}'s$ as follows:
\begin{equation}
{\bf t}_{lk} = {\bf u}_{lk}[1+d\cos{({\bf q}_1\cdot {\bf R}_{lk})}] 
\label{eq:displacement}
\end{equation}
where $l$ is the index of the unit cells inside the ORTHO-IV phase island and $k$ is the atom index in that unit cell. Since the ${\bf u}_{lk}$, as reported in Ref.\cite{A-modulation}, lie in the {\bf a}-{\bf c} plane, the second term of Eqn.(2) represents an additional transverse modulation along the {\bf b}-direction. The second term contributes to the ${\bf q}_1$ shoulders and has small effects on ${\bf q}_0$ peaks. Thus fitted values of ${\bf u}_{lk}$ remain close to those reported in Ref.\ \cite{A-modulation} within the errors introduced by the IFS term included in this work.

To determine the values of $\delta$ and $d$, the experimental data is fitted with the ORTHO-IV model, together with the new lattice modulation as in Eqn.(2). With this model the 2D pattern can be reproduced remarkably well as shown in Fig.\ \ref{fig:data}(b). Fig.\ \ref{fig:fit} gives a better quantitative view of the data fit along important lines indicated in Fig.\ \ref{fig:data}.
Thus, our experimental observation and quantitative analyses establish the existence of {\bf b}-direction lattice modulation with ${\bf q}_1=(0,\delta\approx0.21,0)$ within the ORTHO-IV phase islands in optimally doped YBCO. ${\bf q}_1$ is very close to  ${\bf q}_\text{IC}$, strongly indicating that this modulation is induced by the FS.

The idea that FS nesting can influence short-range order (SRO) and manifest itself in the X-ray diffuse scattering was discussed\cite{moss69} in the study of order-disorder transitions in binary alloys such as Cu-Au and was used subsequently in  investigations\cite{cmdm} of SRO correlated micro-domain structures within a disordered matrix. In the context of the ORTHO-IV phase any one of the four permutations of $<1110>$ can form as anti-phase (AP) domains with periodicity along the {\bf b} axis determined by ${\bf q}_1$ giving rise to symmetric shoulders of ${\bf q}_0$, along K axis. However, structure-factor calculations based on various combinations ({\it e.g.} two $<0111>$ and three $<1110>$; as illustrated in Fig.\ \ref{APD}) assuming no amplitude modulations of the ORTHO-IV displacement pattern\cite{A-modulation} did not yield relative intensities of ${\bf q}_0$ and ${\bf q}_1$ satellites consistent with the data. Although we can not rule out a more subtle arrangements of AP domains, a simple transverse modulation of a mono-domain ORTHO-IV pattern, as modeled in Eqn. (2), adequately accounts for the data.

It is important to note that the shoulders get contributions from all the transversely displaced atoms in the ORTHO-IV islands (predominantly the Ba atoms), rather than from the chains alone. In our view,  nested `ridges' (or `sheets') of the FS simply induce a ${\bf q}_1$ amplitude modulation of the ORTHO-IV displacement pattern with a concomitant presence of a CDW on the CuO chains. Since the ORTHO-IV displacement pattern is a {\it static} relaxation of the lattice\cite{DdF02}, we expect the ${\bf q}_1$ modulation to be static as well. The formation of ${\bf q}_1$ modulations only inside the ORTHO-IV phase islands rather than throughout the average lattice seems to disagree with surface-sensitive STM experiments\cite{STM-Edwards,STM-Maki} which have reported well-patterned electronic corrugations in regions away from oxygen vacancies. This discrepancy may be due to energetics of CDW formation on chain-terminated surfaces as opposed to that in the bulk. The LDA calculations\cite{bandcalcu} indicate that the CuO chains in the bulk are on the verge of a CDW instability due to FS nesting, but whether such a CDW can form or not depends on energy gain of carriers relative to the energy costs due to the lattice deformations. The lack of any peaks (or rods) in the vicinity of $(m, n, L)\pm{\bf q}_1$ in present study indicates that the average lattice may be too rigid for a CDW to form. Instead, it seems that a static CDW can only be established in favorable regions, such as the cleaved surface with CuO chains to be the first layer, as studied with STM, or in other elastically ``softer'' areas. A lattice softening introduced by oxygen vacancies is suggested by STM observations\cite{STM-Edwards,STM-Derro,STM-Maki} of enhanced CDW amplitudes in their vicinity. X-ray studies\cite{A-modulation,ZI02} have also shown that ORTHO-IV islands are softer than the average lattice with structurally coherent chains, which in principle can favor the CDW formation. Since the displacements are transverse to ${\bf q}_1$ and are three dimensional in nature, a straightforward interpretation based on a canonical longitudinal CDW seems precluded. However, we can speculate that ORTHO-IV islands may act as `pinning' templates\cite{ZI02} for dynamic charge fluctuations\cite{HAM}, calling for further studies of FS effects in YBCO.

In conclusion, in optimally doped YBCO, an incommensurate transverse lattice modulation with wavevector ${\bf q}_1=(0,\sim0.21,0)$ is observed inside the ORTHO-IV phase islands. This modulation is induced by FS nesting associated with 1D CuO chain subsystems and forms inhomogeneously in the lattice. Since the CuO$_2$ planes are also involved in the ${\bf q}_0\pm{\bf q}_1$ modulation, FS effects must affect the planes as well.

Research at UCSD is supported by the U.S. Department of Energy (DOE), Basic Energy Sciences through Grant No. DE-FG02-03ER46084. Use of the APS is supported by the DOE, Office of Science, under Contract No. DE-AC02-06CH11357. At Houston support comes from the State of Texas through the Texas Center for Superconductivity at the University of Houston (TcSUH).

\end{document}